\pdfoutput=1

\documentclass[reprint,
 amsmath,amssymb,
 aps,
]{revtex4-1}

\usepackage{natbib}
\usepackage{graphicx}
\usepackage{hyperref}
\usepackage[title]{appendix}
\usepackage[font={small}]{caption}
\usepackage{enumitem}
\usepackage{float}
\usepackage{url}
\usepackage{bm}

\renewcommand{\l}{\ensuremath{\langle}}
\renewcommand{\r}{\ensuremath{\rangle}}

\newcommand{\lb}{\ensuremath{\left(}}
\newcommand{\rb}{\ensuremath{\right)}}

\newcommand{\lbb}{\ensuremath{\left[}}
\newcommand{\rbb}{\ensuremath{\right]}}

\newcommand{\C}{\ensuremath{A}}
\newcommand{\CC}{\ensuremath{B}}
\newcommand{\D}{\ensuremath{\Delta}}
\newcommand{\m}{\ensuremath{\mu}}
\newcommand{\si}{\ensuremath{G}} 
\newcommand{\s}{\ensuremath{\Sigma}}
\renewcommand{\ss}{\ensuremath{\Gamma}} 
\newcommand{\ssr}{\ensuremath{\epsilon}} 
\newcommand{\ssa}{\ensuremath{\lvert \s \rvert}} 

\newcommand{\de}{\ensuremath{\delta}}
\newcommand{\p}{\ensuremath{\varphi}}

\newcommand{\Pb}{\ensuremath{\Psi}}

\renewcommand{\k}{\ensuremath{\mathbf{k}}}
\renewcommand{\bf}{} 
\newcommand{\hc}[1]{#1^{\dag}}
\newcommand{\co}[1]{#1^{*}} 
\newcommand{\g}[3]{\co{G}_{#2,#1} G_{#3,#1}}
\newcommand{\be}{\begin{equation}}
\newcommand{\ee}{\end{equation}}
\newcommand{\fd}{\ensuremath{\operatorname{N_F}}}
\newcommand{\intf}{\ensuremath{\int_{-\infty}^{\infty}}}
\newcommand{\im}{\ensuremath{\operatorname{Im}}}
\newcommand{\al}[1]{\begin{align*}#1\end{align*}}
\newcommand{\all}[1]{\begin{align}#1\end{align}}
\newcommand{\alll}[1]{\be \begin{aligned}#1\end{aligned} \ee}
\newcommand{\sumk}{\ensuremath{\sum\limits_{\k}}}

\begin{document}


\title{Conductance of a Finite Kitaev Chain}

\author{R.J. Doornenbal}
\email{R.J.Doornenbal@uu.nl}
\author{G. Skantzaris}
\author{H.T.C. Stoof}
\affiliation{
Institute for Theoretical Physics and Center for Extreme Matter and Emergent Phenomena,\\
Utrecht University, Leuvenlaan 4, 3584 CE Utrecht, The Netherlands
}

\date{\today}

\begin{abstract}
We present a stochastic formulation of the Keldysh theory to calculate the conductance of a finite Kitaev chain coupled to two electron reservoirs. We study the dependence of the conductance on the number of sites in the chain and find that only for sufficiently long chains and in the regime that the chain is a topological superconducter the conductance at both ends tends to the universal value $2e^2/h$, as expected on the basis of the contact resistance of a single conducting channel provided by the Majorana zero mode. In this topologically nontrivial case we find an exponential decay of the current inside the chain and a simple analytical expression for the decay length. Finally, we also study the differential conductance at nonzero bias and the full current-voltage curves. We find a nonmonotonic behavior of the maximal current through the Kitaev chain as a function of the coupling strength with the reservoirs.
\end{abstract}


\maketitle

\section{Introduction}

Majorana fermions were first considered in high-energy physics as hypothetical elementary particles with the defining property that these fermions are their own antiparticles, in the same spirit that the photon is its own antiparticle in the bosonic case. As elementary particles Majorana fermions have not yet been proven to exist, but condensed-matter quasiparticles with similar characteristics are theoretically known to appear under suitable circumstances \cite{Kitaev2000}. These Majorana quasiparticles are presently of great interest for their possible applications in quantum computing as it is predicted that a pair of spatially separated Majoranas can be used to construct a topologically protected qubit which is insensitive to decoherence.
In recent years, compelling experimental evidence of Majorana fermions has been found in semiconductor nanowires \cite{Mourik2012,Das2012,Deng2012} following the specific theoretical proposals in Refs.\ \cite{Sau2010,Lutchyn2010,Oreg2010}. In these experiments the signature of the Majorana fermions came in the form of a zero-bias peak in the measurements of the differential conductance of the nanowires that was mounted on a superconducting substrate and placed in a magnetic field.

In principle, the magnetic field, together with the spin-orbit coupling in the semiconductor nanowire, plays a crucial role in the realization of the above-mentioned topological superconductor and its Majorana zero modes. However, there exists a more simple model of a one-dimensional chain that nevertheless captures the essential features of topological superconductivity and the emergence of Majorana fermions. This is the famous Kitaev model \citep{Kitaev2000}, whose resistance properties we aim to study in this paper.

In view of their potential for the creation of topological qubits, there has already been considerable effort to study analytically and numerically the conductance of topological superconductors in which Majorana fermions can emerge, including the so-called Kitaev chain \cite{Law2009,Flensberg2010,Bolech2007, Lutchyn2013}.
However, most studies have focused on the thermodynamic limit of infinitely long chains. Moreover, in case of the Kitaev chain, they were restricted to the special parameter values where the Kitaev model is exactly solvable and the Majoranas are perfectly localized at the first and last site of the chain. But it is also experimentally relevant to study chains of a finite length. This is because the nanowires discussed by Mourik {\it et al}. \citep{Mourik2012} are thought to be segmented by disorder into a number of smaller coherent chains. In the present paper, we therefore investigate the conductance of the Kitaev chain coupled to leads, focussing especially on finite-size effects and the effects of imperfect localization of the Majorana fermions.
Note that we do not explicitly consider the effect of disorder, which is for instance the topic of recent work in Refs.\ \cite{Fregoso2013,Wang2013,Lobos2014}. The effect of a nearest-neighbor Hubbard interaction on the Kitaev model, as discussed in Ref.  \cite{Thomale2013}, is also not considered here.

An important aspect of the Kitaev model is that the superconducting order parameter is not determined self-consistently from a gap equation, but is just a free parameter of the model. Physically the superconductivity of the chain is therefore induced by a proximity effect of a superconducting substrate that acts as an electron reservoir. As a result there is no charge conservation in the Kitaev chain and the current is, also in a steady-state situation, spatially inhomogeneous. This is possibly also of experimental interest, because it in principle allows for a direct observation of the wavefunction of the Majorana fermion in a semiconductor nanowire mounted on top of a superconductor.

The paper is organized as follows. In Sec.\ II we present a stochastic formulation of the nonequilibrium Keldysh theory for a Kitaev chain that is connected at its end points by a tunnel junction to an electron reservoir and show how it can be used to determine the conductance of a finite Kitaev chain. In Sec.\ III we present the zero-temperature results based on this approach. In particular we determine the conductance in various different areas of the phase diagram of the Kitaev model. We determine the inhomogenous current profile in the chain and also go beyond linear response to consider the full $I-V$ curves and associated differential conductance. We end in Sec.\ IV with our conclusions and an outlook for further investigations.

\section{Keldysh theory for the Kitaev chain}

In this section, we first briefly describe the Kitaev model and show how it can be coupled to two leads at the left and right end points of the chain. We then present the Keldysh theory for the steady-state solution with a voltage difference over the chain and use this solution to appropriately define the conductance in this case.

\subsection{Kitaev model}

We start by considering a tight-binding chain of $N$ sites. Electrons can only hop to nearest-neighbor sites and are assumed to be spin polarized. Neighboring electrons can also form Cooper pairs, which is described by a $p$-wave-pairing term $\D^* a_{j+1} a_j + \D a^{\dagger}_{j} a^{\dagger}_{j+1}$ for each pair of sites in the Hamiltonian. Here $\D$ is the superconducting order parameter and $a_j$ is the annihilation operator for an electron on site $j=1, \dots , N$. The grand-canonical Hamiltonian, first proposed in the context of Majorana fermions by Kitaev \cite{Kitaev2000}, becomes then
\alll{
H = & -  t \sum \limits_{j=1}^{N-1} ( \hc{a}_{j+1} a_j +  \hc{a}_j a_{j+1})
      - \m  \sum \limits_{j=1}^N  \hc{a}_j a_j \\
    & + \sum \limits_{j=1}^{N-1} ( \D^* a_{j+1} a_j + \D \hc{a}_{j} \hc{a}_{j+1}) ,
}
where $t$ is the hopping amplitude and $\m$ is the chemical potential.
By rescaling the creation and annihilation operators by a phase factor, we can change the phase of the anomalous terms proportional to \D. We may thus assume without loss of generality that \D\ is real and nonnegative. Moreover, from now on we measure all energies in units of $t$ and use units such that $\hbar = e = 1$. To convert a dimensionless expression for conductance to SI units, we thus have to multiply the result by $e^2/\hbar$. In particular, a conductance of $1/\pi$ is equal to $2e^2/h$ in SI units.

The Kitaev model with $\D > 0$ is known to exhibit a topologically nontrivial superconducting phase for $|\m| < 2$, and a topologically trivial superconducting phase for $|\m| > 2$. In the topological superconductor phase, there are localized Majorana zero modes at each end, with an exponentially decaying wavefunction. At the special, and exactly solvable, point $\D = 1$ and $\m = 0$, these Majoranas are even confined to the first and last lattice sites, without any wavefunction overlap \cite{Alicea2012}. For future reference we note that the Hamiltonian in Eq.~(1) may be written in matrix form as
\alll{
H = \frac{1}{2} ( \hc{a}_1, ... , \hc{a}_N, a_1, ... a_N ) \cdot {\bf K}  \cdot \begin{pmatrix}
  a_1 \\
  \vdots \\
  a_N \\
  \hc{a}_1 \\
  \vdots \\
  \hc{a}_N
 \end{pmatrix},
}
where the $2N \times 2N$ matrix ${\bf K}$ is given by
\alll{
{\bf K} =
 \begin{pmatrix}
  {\bf K}_0 & {\bf D} \\
  -{\bf D} & -{\bf K}_0
 \end{pmatrix}
}
in terms of the two $N \times N$ matrices
\alll{
{\bf K}_0 =
\begin{pmatrix}
-\m & -t & 0 & \cdots & \cdots & 0 \\
-t & -\m & -t & 0 & \cdots & 0 \\
0 & -t & \ddots & \ddots & \ddots & \vdots \\
\vdots & \ddots & \ddots & \ddots & -t & 0 \\
0 & \cdots & 0 & -t & -\m & -t \\
0 & \cdots & \cdots & 0 & -t & -\m \\
\end{pmatrix}
}
and
\alll{
{\bf D} =
\begin{pmatrix}
0 & \D & 0 & \cdots & \cdots & 0 \\
-\D & 0 & \D & 0 & \cdots & 0 \\
0 & -\D & \ddots & \ddots & \ddots & \vdots \\
\vdots & \ddots & \ddots & \ddots & \D & 0 \\
0 & \cdots & 0 & -\D & 0 & \D \\
0 & \cdots & \cdots & 0 & -\D & 0 \\
\end{pmatrix}~,
}
describing the normal tight-binding chain and the $p$-wave pairing, respectively.

\subsection{Connection to leads}

We next connect the chain at the left and right ends to reservoirs with an ideal electron gas. The reservoirs are kept at a different chemical potential $\m_L$ and $\m_R$, respectively, but are otherwise identical. Without loss of generality, we assume that $\m_L \geq \mu \geq \mu_R$, so that the electron current on average flows from left to right. Letting $a_{\k,L}$ denote the annihilation operator of an electron in the left reservoir with wave vector $\k$ and energy $\epsilon_{\bf k}$, the reservoirs can be described by adding terms to the Kitaev Hamiltonian of the form
\be
H_L = \sumk (\epsilon_\k-\mu_L) \hc{a_{\k,L}}a_{\k,L} - \sumk t_\k (\hc{a_{\k,L}}a_1 + \hc{a_1}a_{\k,L})~,
\ee
and a similar expression for the right reservoir. Here $t_\k$ characterizes the tunneling strength between the left reservoir and the first site of the chain and the right reservoir and the last site of the chain. It will be proportional to $1/\sqrt{V}$, where $V$ is the volume of the reservoirs. Eliminating the reservoirs, these couplings induce self-energy terms for the electrons in the chain which may be calculated for example by using second-order perturbation theory and are equal to
\be
\s_1(E) = \sumk \frac{|\l \k |H_L| 1 \r |^2}{E-\epsilon_{\k,L} + i 0} = \sumk \frac{t_\k^2}{E-\epsilon_{\k,L}+ i 0}~,
\ee
where we introduced the notation $\epsilon_{\k,L} = \epsilon_{\k} -\mu_L$. A similar expression holds for $\s_N(E)$. Denoting the principle value part by $\mathcal{P}$ we can now write
\alll{
\s_1(E) & = \sumk t_\k^2 \frac{\mathcal{P}}{E - \epsilon_{\k,L}} - i \pi \sumk t_\k^2 \, \de(E-\epsilon_{\k,L}) \\      & \equiv \ssr_1(E) - i \ss(E)/2.
}
If the density of states of the reservoir and the tunneling parameter $t_\k$ are approximately constant near the energies accessible to the chain, then $\s_1(E)$ may be taken to be the constant $\s_1(0)$. The real part $\ssr_1$ of $\s_1$ induces a shift in the energy at the left endpoint of the chain. We will set it to zero for simplicity unless otherwise noted. All results can in principle be extended to the case where the real part is nonzero, although the final expressions become more complicated. We also always take the self-energies from both reservoirs to be equal, neglecting any differences caused by, for example, the differences in chemical potential. We therefore simply write $\s_1 = \s_N = \s = \ssr - i \ss/2$. Note that in this approximation $\ss$ is simply proportional to the density of states of the reservoir at the Fermi level. In analogy with the matrix $K_{\alpha,\beta}$, we finally define also the $2N \times 2N$ self-energy matrix $\s_{\alpha,\beta}$ by $\s_{1,1} = \s_{N,N} = \s, \s_{N+1,N+1} = \s_{2N,2N} = -\co{\s}$, and all other entries zero.

It can be shown using the Keldysh formalism \cite{stoof1998} that the influence of the reservoirs may be entirely incorporated into the theory by the above `retarded' self-energy matrix for the Kitaev chain and additional `noise' sources, which are nonzero only on the first and last sites and physically represent the shot noise of hopping of electrons to and from the reservoirs. We do not derive this stochastic formulation of the Keldysh theory but give here only a phenomenological derivation which is analogous to the description of classical Brownian motion. For generality, we will only set $\s_{\alpha,\beta}(E)$ to a constant at the end of the calculation.

In the stochastic formulation the second-quantization operators $a_j$ are replaced by (anticommuting) complex fields $\p_j$ that are subject to fluctuations described by a Langevin equation and obey $\l \p_j \r = \l a_j \r$ after averaging over the noise. Introducing as in Eq.\ (2) the Nambu-space-like vector $\Pb$ with $2N$ components $(\p_1, ..., \p_N, \co{\p}_1, ...,\co{\p}_N)$, the Langevin equation may in this case thus be written as
\be
i \frac{d}{dt} \Pb_\alpha(t) - K_{\alpha,\beta} \Pb_\beta(t) - \int \limits_{-\infty}^{\infty} dt'\, \s_{\alpha,\beta}(t-t') \Pb_\beta(t') = \eta_\alpha(t)~,
\ee
where summation over repeated indices is assumed. Moreover, $\s_{\alpha,\beta}(t)$ is the Fourier transform of $\s_{\alpha,\beta}(E)$ introduced above. The vector $\eta_\alpha$ of length $2N$ is again given by $(\eta_1, ..., \eta_N, \co{\eta}_1, ...,\co{\eta}_N)$ and represents the noise induced by the reservoirs. Clearly the only nonzero components of $\eta_j$ are for $j=1,N$.

Fourier transforming the Langevin equation in Eq.~(9) yields
\alll{
\ \eta_\alpha(E) &= \lbb E\de_{\alpha,\beta} - K_{\alpha,\beta} - \s_{\alpha,\beta}(E) \rbb \Pb_\beta(E) \\ &\equiv G^{-1}_{\alpha,\beta}(E)\Pb_\beta(E)~,
}
where we defined the (frequency space) inverse retarded Green's function matrix ${\bf G}^{-1}(E)$. This yields \all{\label{eqn:expecpsi} \l {\Pb}^*_\alpha(E) \Pb_\beta(E') \r = \co{G}_{\alpha,\alpha'}(E) G_{\beta,\beta'}(E') \l \co{\eta}_{\alpha'}(E) \eta_{\beta'}(E') \r~.}
Since the noise amplitudes are uncorrelated at different sites and also at different energies for steady-state applications, we have that
\alll{
&\l \co{\eta}_j(E) \eta_{j'}(E') \r = A_j(E) \delta_{j,j'} \de(E - E'),\\
&\l {\eta}_j(E) \co{\eta}_{j'}(E') \r = B_j(E) \delta_{j,j'} \de(E - E'),
}
where the functions $A_j(E)$ and $B_j(E)$ are only nonzero for $j=1,N$ and determine the strength of the noise at these sites. These functions are in the microscopic Keldysh theory determined from the so-called `lesser' and `greater' self-energies, respectively, but can also be easily obtained phenomenologically from the fluctuation-dissipation theorem in the following manner.

Because the reservoirs are independent of each other, the noise correlation can only be nonzero for the first and last sites of the chain. Hence it suffices to find the noise correlation function induced by a single reservoir at chemical potential $\m$ coupled to a single site. For the single site, dropping the redundant indices and using for a moment canonical energies, we readily find $G(E) = 1/(E-\s(E))$, hence
\be \l \co{\p}(E) \p(E) \r = \dfrac{1}{|E-\s(E)|^2} \l \co{\eta}(E) \eta(E) \r. \ee
The fluctuation-dissipation theorem \cite{noneq} states that in equilibrium
\be \l \hc{a}(E) a(E) \r = -2 \fd(E-\m) \im G(E) \ee
and
\be \l a(E) \hc{a}(E)  \r = -2  \lbb 1-\fd(E-\m) \rbb \im G(E)~.\ee
Here $\fd(x) = \left[1 + \exp(\beta x) \right]^{-1}$ is the Fermi-Dirac distribution.
Demanding that $\l \co{\p}(E) \p(E) \r = \l \hc{a}(E) a(E) \r$ and $\l \p(E) \co{\p}(E) \r = \l a(E) \hc{a}(E) \r$, we can combine these equations to obtain the desired result
\be \C(E) = -2 \im \s(E) \fd(E-\m) \ee
and \be \CC(E) = -2 \im \s(E) \lbb 1 - \fd(E-\m) \rbb. \ee

Returning to the Kitaev chain, we thus conclude that
\alll{\l \co{\eta}_\alpha(E) \eta_\beta(E') \r & = \ss \de_{\alpha,\beta} \de(E - E') \\
&\times
\begin{cases}
  \fd(E+\mu-\m_L) & \mbox{if } \alpha = 1  \\
  \fd(E+\mu-\m_R) & \mbox{if } \alpha = N \\
1-\fd(E+\mu-\m_L) & \mbox{if } \alpha = N + 1  \\
1-\fd(E+\mu-\m_R) & \mbox{if } \alpha = 2N \\
            0 & \mbox{else}
\end{cases} ,
}
where we incorporated the use of a grand-canonical Hamiltonian with the chemical potential $\mu$ by replacing in Eqs.\ (16) and (17) $E$ by $E + \m$.

\subsection{Current and conductance}

In this section, we show how we can define and calculate the conductance of the Kitaev chain using Eqs.\ (11) and (18), which allow us to calculate all the expectation values of the form $\l \hc{a}_j a_{j'} \r$. In the normal state of the chain, i.e., $\Delta = 0$, the current through the system is spatially constant in a steady-state situation and the conductance can be obtained by calculating the current anywhere inside the chain and differentiating with respect to the voltage difference over the chain. In the superconducting state the current is however spatially inhomogeneous and we have to make sure that we are properly calculating the current coming out or going into the reservoirs to determine the conductance. Interestingly, in this case we actually only require the presence of a single reservoir at one of the ends of the chain to obtain a steady-state situation. Physically this comes about because the superconducting substrate that induces the superconductivity in the chain by the proximity effect provides the other reservoir of electrons.

\subsubsection{\label{sec:current} Current operator and charge conservation}

To determine the local current we consider the electron-number operator $n_j = \hc{a}_j a_j$ at site $j$, which we assume for now to have neighbors to both its left and right. We then have
\be
\begin{split}
\frac{d n_j }{dt} &= i [H,n_j] \\
 &= - i \D \lb a_{j+1} a_{j} - \hc{ a}_j  \hc{ a }_{j+1} \rb\
 - i \D \lb a_{j} a_{j-1} - \hc{ a}_{j-1}  \hc{ a }_{j}  \rb  \\
 &+ i \lb \hc{a}_{j+1} a_{j}  - \hc{a}_{j} a_{j+1} \rb - i \lb \hc{a}_{j} a_{j-1} - \hc{a}_{j-1}a_{j} \rb~.
\end{split}
\ee
In this expression we recognize four terms in the right-hand side. The first two terms represents the loss of electrons due to the formation of Cooper pairs with the site to the right and with the site to the left, respectively. The third term represents the loss of electrons due to electrons hopping to the site to the right and similarly the fourth term represents the gain of electrons due to electrons hopping from the site to the left. If the neighbor to the left and/or right is absent, the corresponding terms are of course also absent.
Identifying the latter two terms with the well-known tight-binding particle current between sites \cite{Mahan}, the average particle current flowing from site $j$ to site $j+1$ becomes $J_{j,j+1} = -i  ( \l\co{\p}_{j+1} \p_{j}\r  - \l \co{\p}_{j} \p_{j+1} \r )$. The terms proportional to $\D$ are identified with the flow of electrons from site $j$ to and from the superconductor and are ultimately responsible for the inhomogeneous current profile in a superconducting Kitaev chain.

In particular, this also allows us to calculate the current flowing into the chain from a reservoir. By charge conservation, the current flowing into the first site must equal the current $J_{1,2}$, plus the net flow of electrons from site $1$ to the superconductor. We have explicitly checked that this relation is indeed satisfied by adding a non-superconducting site $j = 0$ between the reservoirs and the chain, i.e., a site that is only coupled to the chain by hopping and not via the anomalous terms proportional to \D. The current $J_{0,1}$ was explicitly calculated and was indeed found to be identically equal to the current $J_{1,2}$ plus the Cooper pair current from sites 1 and 2.

\subsubsection{Conductance}

In case $\D > 0$, uniquely defining the conductance of the chain becomes somewhat subtle because the current flowing between each pair of sites is not a constant. The currents may also depend nontrivially on how the total drop in chemical potential is distributed over the reservoirs. Therefore we first define the left and right differential conductances matrices $\si_{j,j',L}$ and $\si_{j,j',R}$ by
\begin{align}
\si_{j,j',L} &= \frac{\partial J_{j,j'}}{\partial \m_L},\\
\si_{j,j',R} &= - \frac{\partial J_{j,j'}}{\partial \m_R}.
\end{align}
Here $J_{j,j'} \equiv -i ( \l\co{\p}_{j'} \p_{j}\r  - \l \co{\p}_{j} \p_{j'} \r )$.
It should not lead to confusion that the Green's functions and conductances are both denoted by $G$, because the conductances are always accompanied by the subscript $L$ or $R$ in the following.

We are also interested in the current flowing from the reservoirs into the chain. Let $J$ be the average current flowing from the left reservoir into site 1. We now define
\begin{align}
\si_{L} &= \frac{\partial J}{\partial \m_L},\\
\si_{R} &= - \frac{\partial J}{\partial \m_R}.
\end{align}
We could similarly define conductances related to the current flowing out of the chain from site $N$, but we restricted ourselves to the incoming current. The symmetry of the problem then allows us to also draw conclusions about the right reservoir.

It follows from the definition of $\Pb_\alpha$ that
$ \co{ \p}_j  \co{ \p }_{j'} - \p_{j'} \p_{j} =  \co{ \Pb}_{j} \Pb_{j'+N}  - \co{ \Pb}_{j'+N} \Pb_{j}$.
To obtain the current at any link we therefore only need to calculate equal-time expectation values of the form $i( \l \co{\Pb}_\alpha \Pb_\beta \r - \l \co{\Pb}_\beta \Pb_\alpha \r )$.
Using Eq. (\ref{eqn:expecpsi}) for $\l \co{\Pb}_\alpha \Pb_\beta \r$ yields
\alll{
&i \lb \l \co{\Pb}_\alpha \Pb_\beta \r - \l \co{\Pb}_\beta \Pb_\alpha \r \rb
\\\label{eqn:equicurrent}&= -\frac{\ss}{\pi} \intf dE\, \Big\{  \im \lb \g{N+1}{\alpha}{\beta} + \g{2N}{\alpha}{\beta}  \rb
\\&+ \im \lb \g{1}{\alpha}{\beta} - \g{N+1}{\alpha}{\beta}\rb  \fd \lbb E + \m - \m_L \rbb
\\&+  \im \lb \g{N}{\alpha}{\beta} - \g{2N}{\alpha}{\beta} \rb  \fd \lbb E + \m - \m_R \rbb \Big\}.
}
Here it must be remembered that the Green's function matrix ${\bf G}(E)$ also depends on $E$. Moreover, note that the first term in the right-hand side of this expression is independent of $\mu_L, \mu_R$. This term represents the current flowing through the system when it is in equilibrium with the reservoirs. Symmetry dictates that it must vanish identically. While we were unable to prove this rigorously for general parameter values, we numerically found that it was indeed always the case.

In the zero-temperature limit, the Fermi-Dirac distribution function tends to a step function, and we obtain
\alll{
\si_{j,j',L} &= \frac{\ss}{\pi} \im \lbb \left.(\g{1}{j}{j'} - \g{N+1}{j}{j'}) \right|_{E = \m_L - \m} \rbb, \\
\si_{j,j',R} &= -\frac{\ss}{\pi} \im \lbb \left.(\g{N}{j}{j'} - \g{2N}{j}{j'}) \right|_{E = \m_R - \m} \rbb.
}
Similar expressions hold for $G_L$ and $G_R$, but including some correction terms proportional to $\D$ as argued in Sec. \ref{sec:current}. In the sequel, we always consider the zero-bias limit $\m_L = \m = \m_R$ of these expressions unless otherwise indicated.

\section{Results}

After this introduction to the finite Kitaev chain and how to determine its conductance, we present in this section our results for various regions in the phase diagram. We start the discussion by considering the most studied and exactly solvable case $\D =1$, implying $\D = t$ in dimensionful units. After that we also consider different values of the superfluid order parameter $\Delta$, including also the normal state with $\D=0$. In particular the particle-hole symmetric case with $\mu =0$ allows for analytical analysis and is considered separately. All our calculations are performed at zero temperature, although the general Keldysh theory presented above also applies to nonzero temperatures.

\subsection{The case $\D = 1$}
\label{sec:delta0}

In this subsection we set $\D = 1$ but consider general $\mu$, $\ss$, and $N>3$. Based on calculations with a computer-algebra system, we obtain for the global conductivities of the chain
\alll{
 \si_R &= 0,\\
 \si_L &= \frac{2^{2N} \ss^2}{\pi \lbb \ss^4 \m^{2N-4} + 16 \m^{2N} + \ss^2 \lb 2^{2N} + 8\m^{2N-2} \rb \rbb}~,
}
which leads to the following behavior of $\si_L$ for various values of $N$.

In the thermodynamic limit $N \rightarrow \infty$, we can distinguish two regimes. If $|\m|>2$, the terms involving $\m$ diverge and the conductance vanishes. There are no Majorana zero modes in this case and the superconductor is gapped so electrons cannot tunnel into the chain. If $|\m|<2$, we obtain the universal value $\si_L = 1/\pi$, independent of $\ss$ and $\mu$ in their respective ranges. Restoring the dimensions then yields $\si_L = {2e^2}/{h}$, in accordance with the contact resistance of a single conducting channel \cite{Law2009,Flensberg2010}. The fact that $G_R$ always vanishes exactly is due to the fact that, even if the chain is a topological superconductor, the Majorana zero modes are exactly localized at the first and last site of the chain. As a result, a change of the voltage of the right reservoir does not affect the current flowing out of the left reservoir into the chain.

\begin{figure}[H]
\includegraphics[width=230px]{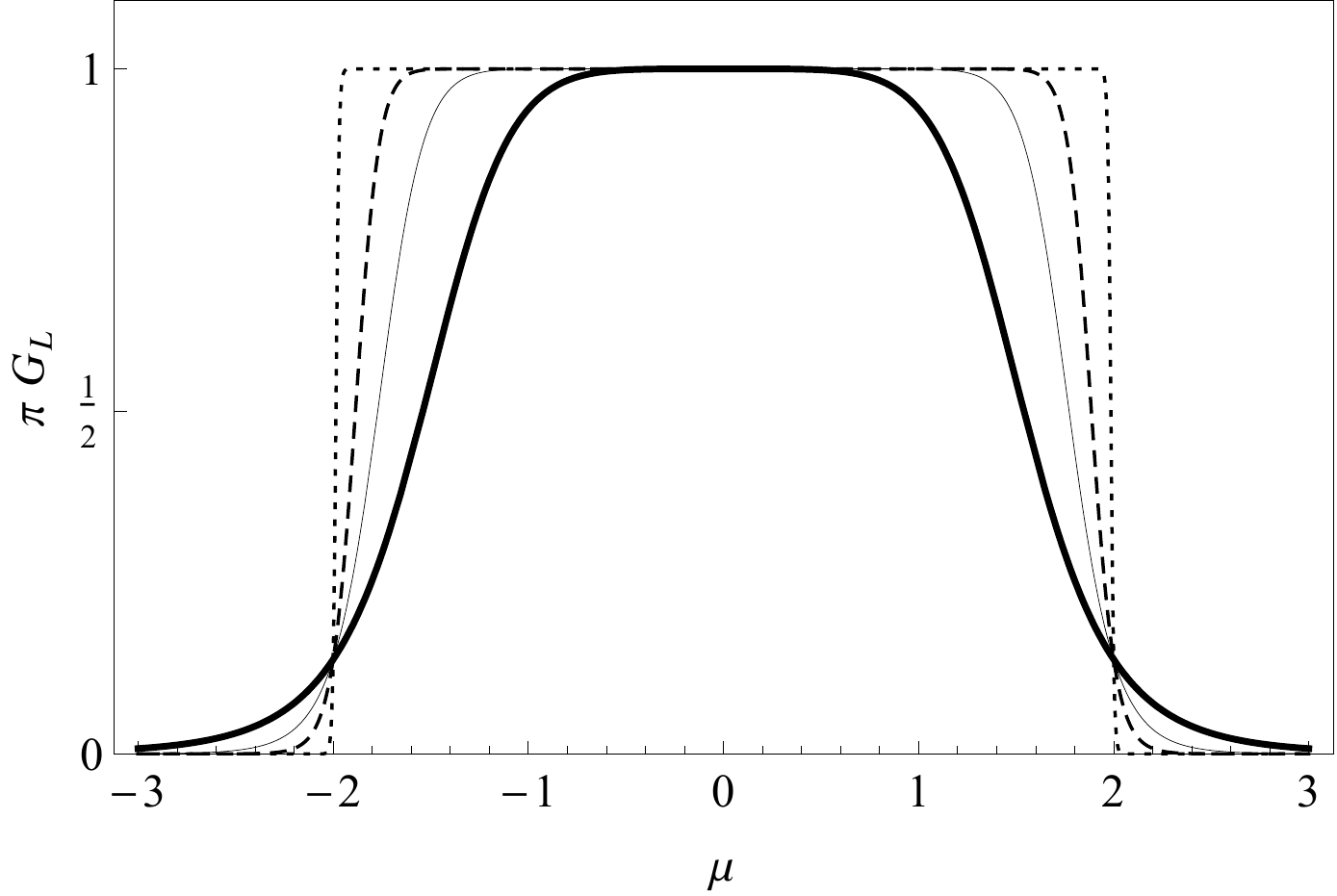}\\
\caption{Conductance of the Kitaev chain at $\D = 1$ and $\ss = 2$, for $N = 4$ (thick solid), $8$ (thin solid), $16$ (dashed), and $128$ (dotted).}
\label{fig:d11}
\end{figure}

\begin{figure}[H]
\includegraphics[width=230px]{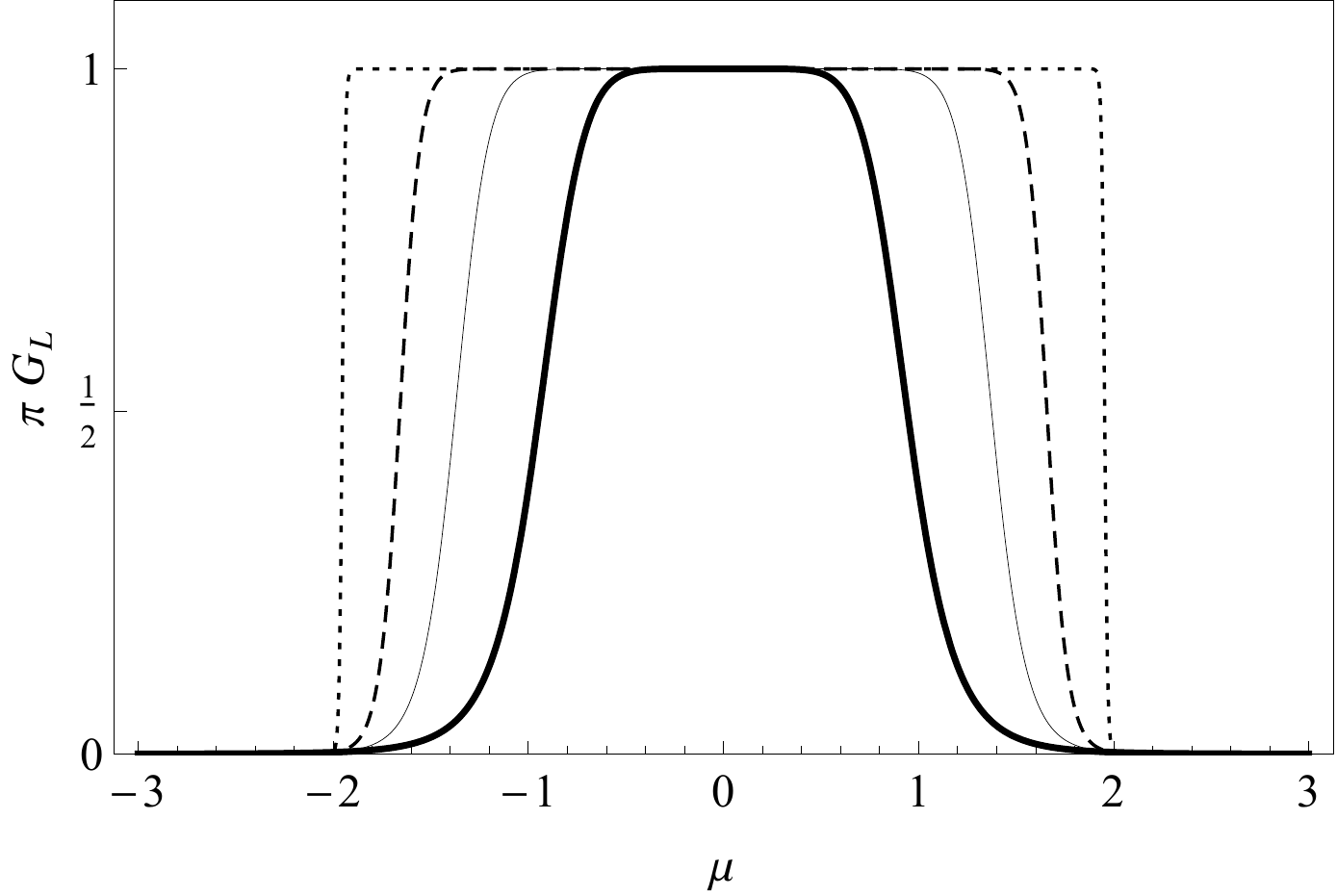}\\
\caption{Conductance of the Kitaev chain at $\D = 1$ and $\ss = 0.2$, for $N = 4$ (thick solid), $8$ (thin solid), $16$ (dashed), and $128$ (dotted).}
\label{fig:d12}
\end{figure}

For finite $N$, the above `box function' of the conductance as a function of the chemical potential $\m$ is smoothed out and the conductance depends in a more complicated manner on \m\ and $\ss$. For $\m = 0$ we still obtain the universal result $\si_L = {2e^2}/{h}$. In Figs.~\ref{fig:d11} and \ref{fig:d12} we have plotted the conductances as a function of $\m$ for $\ss = 2$ and for $\ss = 0.2$, respectively, for various values of $N$. We observe that the value of the coupling to the baths becomes immaterial in the thermodynamic limit but is very important for small system sizes.

We also calculated the local conductances and found that
\begin{align}
\si_{j,j+1,L}  = \begin{cases}
\frac{1}{2}\si_L &\mbox{if $j = 1$},  \\
0 & \mbox{else},
\end{cases}
\end{align}
and similarly
\begin{align}
\si_{j,j+1,R}  = \begin{cases}
\frac{1}{2}\si_L &\mbox{if $j = N-1$},  \\
0 & \mbox{else},
\end{cases}
\end{align}
These results suggest that the Majorana fermions at the endpoints of the chain effectively shield (or short circuit) the chain from the reservoirs, since in linear response no current flows in intermediate links. The fact that $\si_{1,2,L} = \si_L/2$ implies that from the incoming current from the left reservoir exactly half is removed by the formation of Cooper pairs on the first link. The remaining current is then completely transferred into Cooper pairs on the second link so that no current is left over to run through the rest of the chain.

Finally, we give the formulas we obtained for the general case when also the real part $\ssr$ of the self-energy is nonzero, namely
\begin{widetext}
\alll{
\si_L &= \frac{1}{\pi} \frac{2^{2N-4} \ss^2 \lbb
 2 \ssr \ssa^2 \m^{2n-3} + 2 \ssr  \m^{2n-1}+\ssa^4 \mu ^{2 n-4}+2 \ssa^2 \mu ^{2 n-2}+\mu ^{2 n}+2^{2 n-2} \ssa^2 \rbb}{\lbb (2^{N-2} + \m^{2N-4}) \ssa^2 + 2(\ss^2/4 - \ssr^2)\m^{2N-2} +  \m^{2N} \rbb^2 - 2^{2N} \ssr^2 \lbb \ssa^2 \m^{N-2} - \m^N \rbb ^2},\\
\si_R &=\frac{1}{\pi} \frac{2^{2N-3} \m^{2N-3} \ss^2 \ssr \lbb (\ss^2/4 + (\ssr+\m)^2 \rbb}{\lbb (2^{N-2} + \m^{2N-4}) \ssa^2 + 2(\ss^2/4 - \ssr^2)\m^{2N-2} +  \m^{2N} \rbb^2 - 2^{2N} \ssr^2 \lbb \ssa^2 \m^{N-2} - \m^N \rbb ^2}.
}
\end{widetext}
In the limit $\m = 0$, or alternatively in the limit $N \rightarrow \infty$ with $\lvert \m \rvert < 2$, the expression for $G_L$ reduces to \alll{G_L = \frac{1}{\pi} \frac{\ss^2}{\ss^2+4\ssr^2},}
which agrees with the result found in Ref.~\citep{Flensberg2010} for the tunneling conductance into an isolated Majorana state.

\subsection{The case $\mu = 0$}
\label{sec:generald}

We have also obtained closed-form expressions for the conductance in the particle-hole symmetric case $\m = 0$ but for general values of $\D$. They can be elegantly expressed in terms of the polynomials
\be
P_n = \frac{1}{2} \lbb (1+\D)^n + (1-\D)^n \rbb~.
\ee
Furthermore, the results depend on whether $N$ is odd or even. For odd $N$ we have
\alll{
\label{eqn:dodd}
\si_L &= \frac{1}{2\pi} \frac{P_{2N-2}}{P_{N-1}^2}, \\
\si_R &= \frac{1}{2\pi} \frac{(1-\D^2)^{N-1}}{P_{N-1}^2}
}
and for even $N$
\alll{
\label{eqn:deven}
\si_L &= \frac{1}{\pi} \frac{8 \ss^2  P_{2N-2} }{16(1-\D^2)^N + \ss^4(1-\D^2)^{N-2} + 8 \ss^2 P_{2N-2}}, \\
\si_R &= \frac{1}{\pi} \frac{8 \ss^2  (1-\D^2)^{N-1} }{16(1-\D^2)^N + \ss^4(1-\D^2)^{N-2} + 8 \ss^2 P_{2N-2}}.
}
We notice that for odd $N$ the conductance $G_L$ is independent of $\ss$. We find that this also holds for even $N$ if $N$ is large enough. Indeed, in the limit $N \rightarrow \infty$ the conductance $G_L$ tends to the universal value of ${1}/{\pi}$, which again corresponds to ${2e^2}/{h}$ in SI units, independent of $\D$ and $\ss$ as long as they are both nonzero.

\begin{figure}[H]
\includegraphics[width=230px]{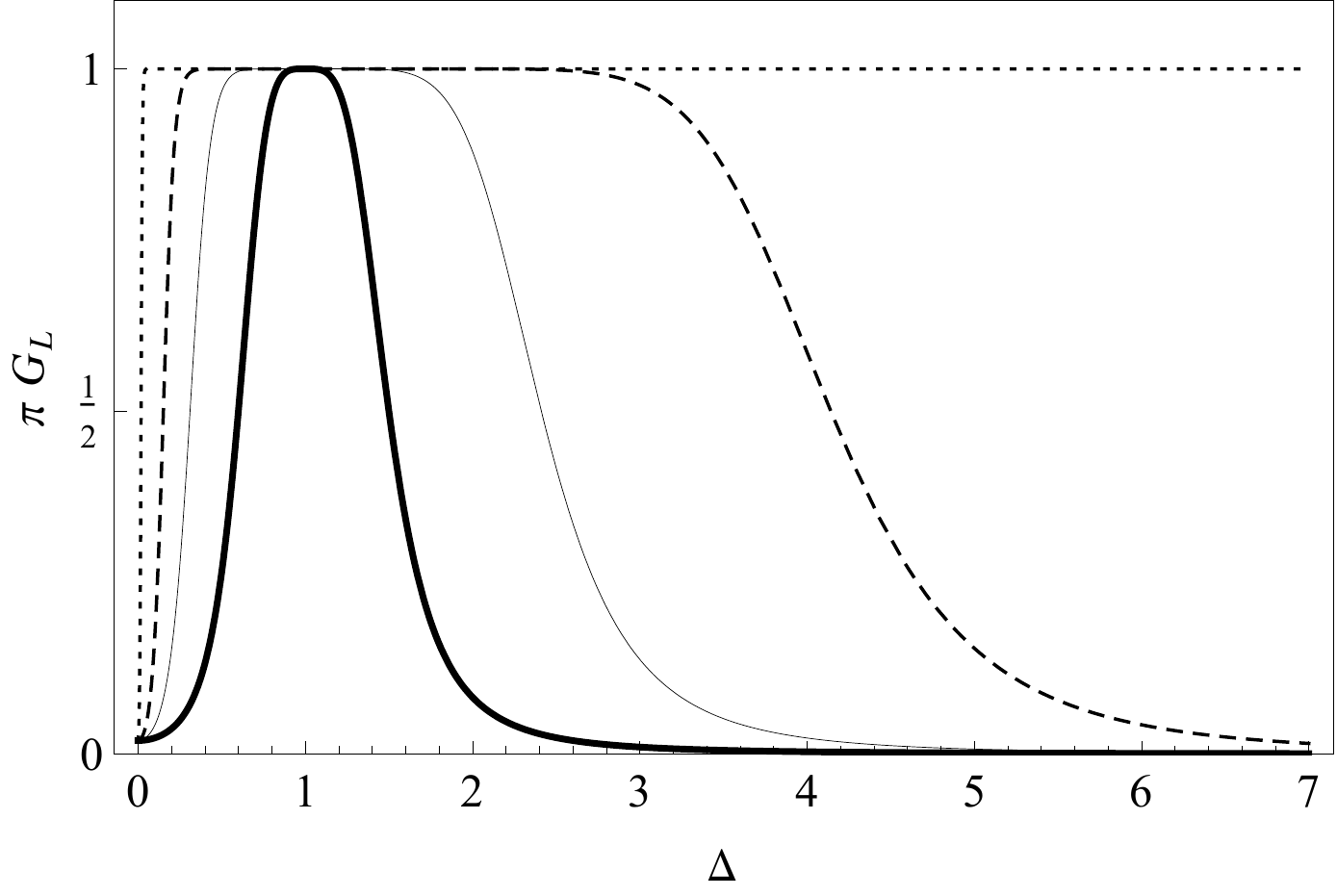}\\
\caption{Left conductance $G_L$ of the Kitaev chain at $\mu = 0$ and $\ss = 0.2$, for $N = 4$ (thick solid), $8$ (thin solid), $16$ (dashed), and $128$ (dotted).}
\label{fig:dvar1}
\end{figure}

\begin{figure}[H]
\includegraphics[width=230px]{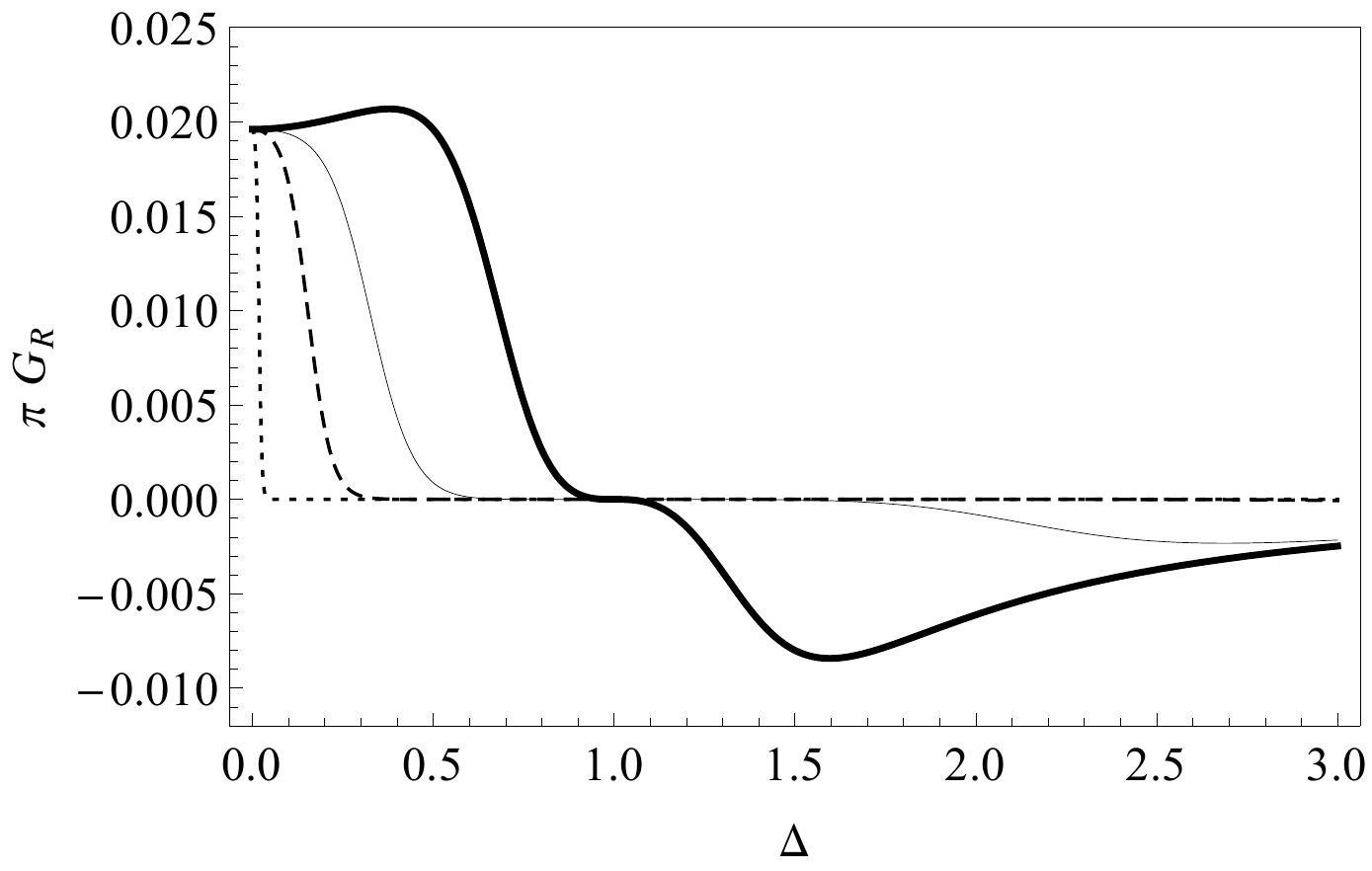}\\
\caption{Right conductance $G_R$ of the Kitaev chain at $\mu = 0$ and $\ss = 0.2$, for $N = 4$ (thick solid), $8$ (thin solid), $16$ (dashed), and $128$ (dotted).}
\label{fig:dvar2}
\end{figure}

The behavior of the `left' and `right' conductances for even $N$ as a function of $\D$ is plotted in Figs.~\ref{fig:dvar1} and \ref{fig:dvar2}, respectively, for various values of $N$. From these figures we can make a number of observations. First, we see that for small system sizes the conductance $\si_L$ exhibits a plateau around $\D = 1$, which broadens with increasing $N$. Second, the conductance $\si_L$ crosses over from this plateau to zero for large values $\D$ at a crossover scale that grows approximately linearly with $N$. Third, both conductances $\si_L$ and $\si_R$ tends to a finite value as $\D \rightarrow 0$ and the chain becomes normal. By taking the appropriate limit in Eq. (\ref{eqn:deven}), we find that the limiting value of the conductance becomes
\be
G_L = \frac{1}{\pi} \frac{\ss^2}{\left(\ss^2/4+1\right)^2} \leq \frac{1}{\pi}~,
\ee
which depends on $\ss$ but not on $N$. Below, in Sec. \ref{sec:ther}, we show that this can be explained by the fact that for even $N$, zero is not an eigenvalue of the Hamiltonian for $\D = 0$ and $\m = 0$. Fourth, the conductance $\si_R$ exhibits a rather complex behavior for small system sizes but becomes simple for large enough values of $N$. Moreover, $\si_R$ becomes negative for $\D > 1$. This turns out to be related to a general feature of the Kitaev chain that some currents `run in reverse' in this regime, as discussed in more detail in Sec. \ref{sec:local} below.

\begin{figure}[H]
\includegraphics[width=230px]{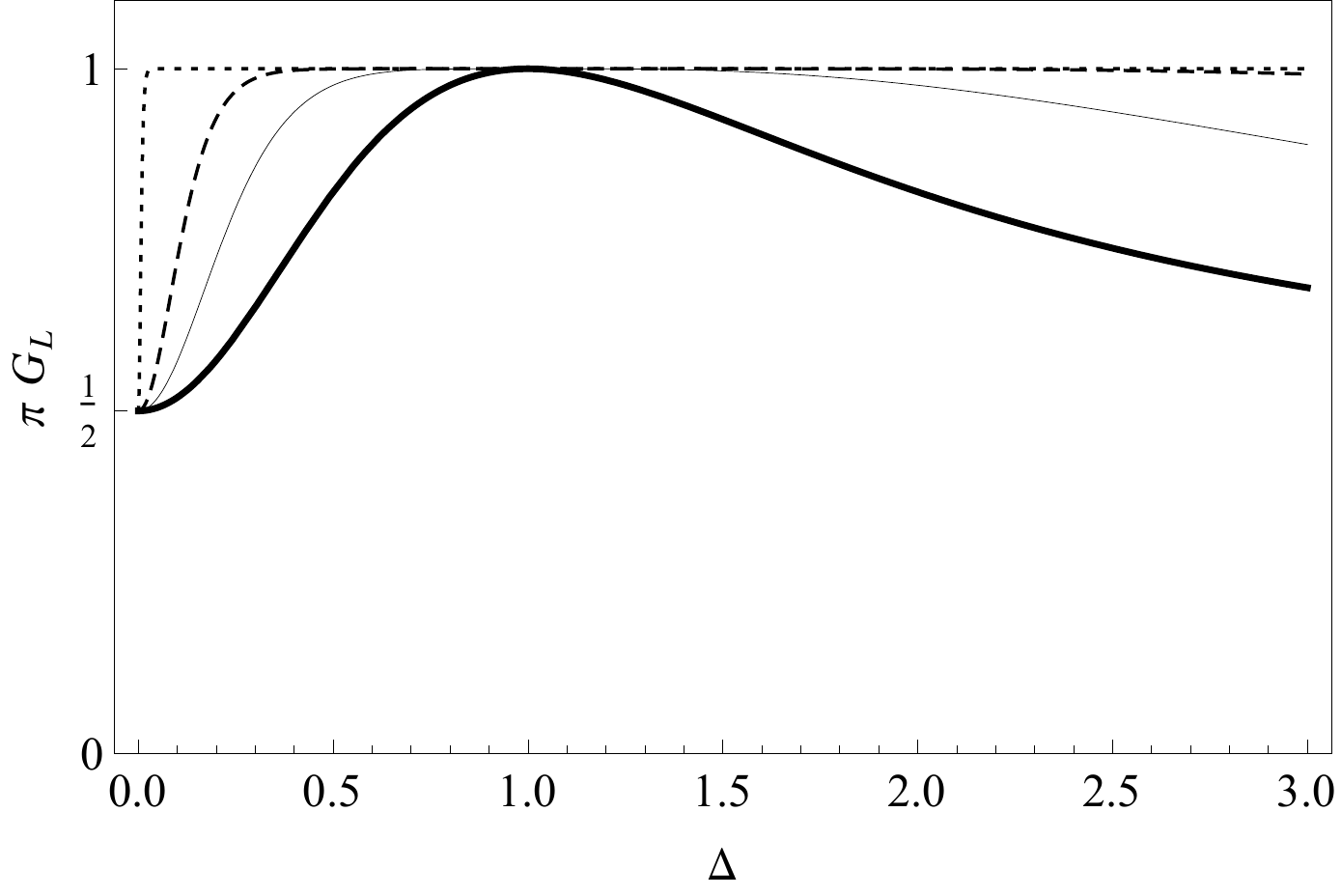}\\
\caption{Left conductance of the Kitaev chain at $\mu = 0$ and arbitrary $\ss$, for $N = 3$ (thick solid), $5$ (thin solid), $9$ (dashed), and $125$ (dotted).}
\label{fig:dodd1}
\end{figure}

Finally, we briefly also look at the case of odd $N$, for which the left conductance is plotted in Fig.~\ref{fig:dodd1}. We find that the behavior of $G_L$ is qualitatively similar to the case of even $N$. Namely, the conductance exhibits a plateau near $\D = 1$ which broadens as $N$ increases. However, an essential difference is that $G_L$ now tends to another universal constant ${1}/{2\pi}$, and not to zero, for both small and large values of $\D$. The graph of $\si_R$ is obtained by reflecting that of $\si_L$ in the line $\si = {1}/{2\pi}$, since $\si_L + \si_R = {1}/{\pi}$ as can be easily proven from Eq.\ (\ref{eqn:dodd})).
In general, we see that as long as $\D > 0$, the conductances tend to the $\D = 1$ values as $N$ tends to infinity. This is consistent with the physical picture of the Majorana fermions becoming localized at the edges of the chain so that if their separation is large enough, the overlap of the wavefunctions becomes negligible even if the localization is not perfect.

\subsection{Fabry-P\'erot resonances}
\label{sec:ther}

\begin{figure}[H]
\includegraphics[width=230px]{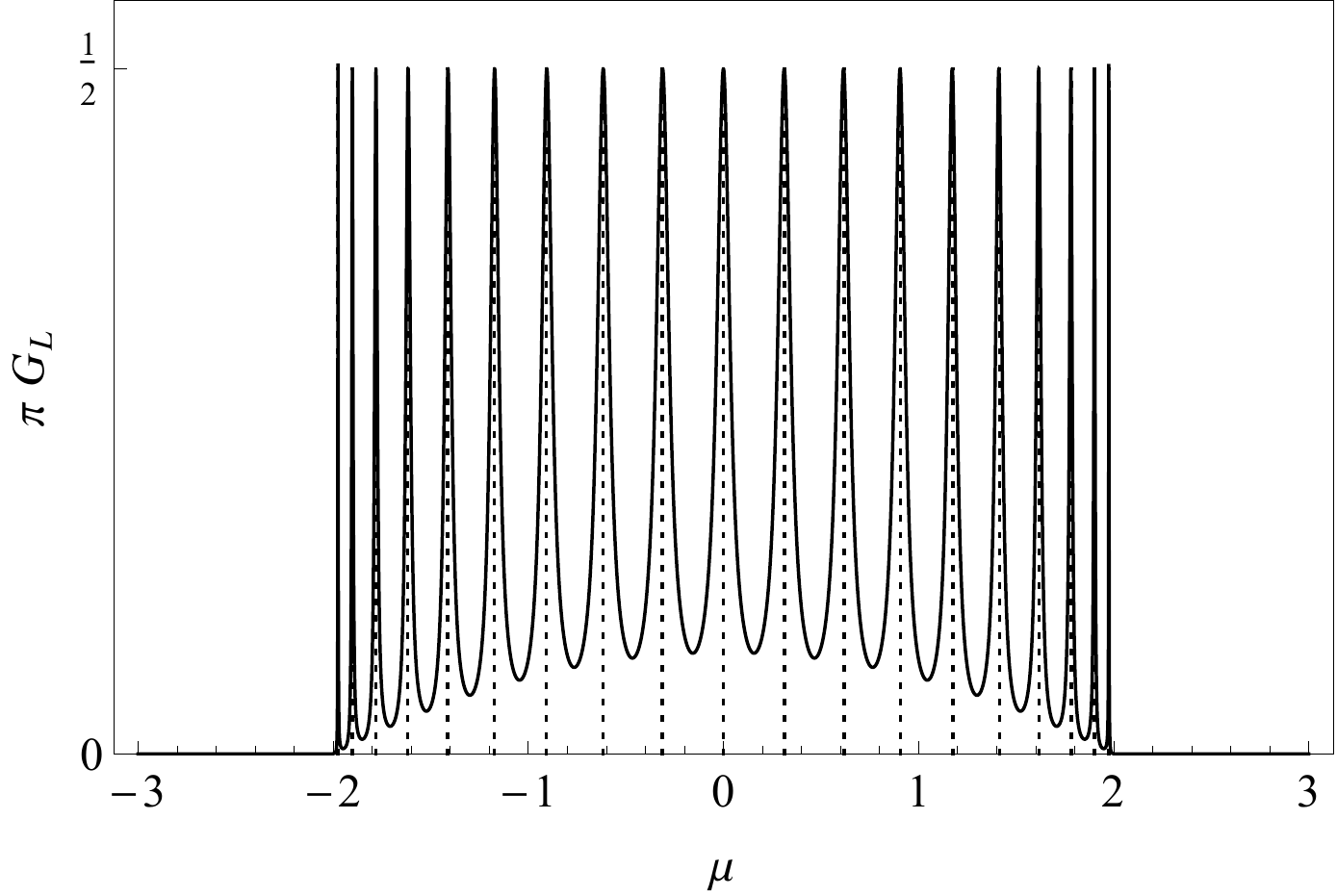}\\
\caption{Conductance of the Kitaev chain as a function of $\m$ for $\ss = 0.4$, $\D = 0$, and $N = 19$. The eigenvalues of the canonical Hamiltonian with $\D = 0$ are shown as dotted lines.}
\label{fig:eig}
\end{figure}

\begin{figure}[H]
\includegraphics[width=230px]{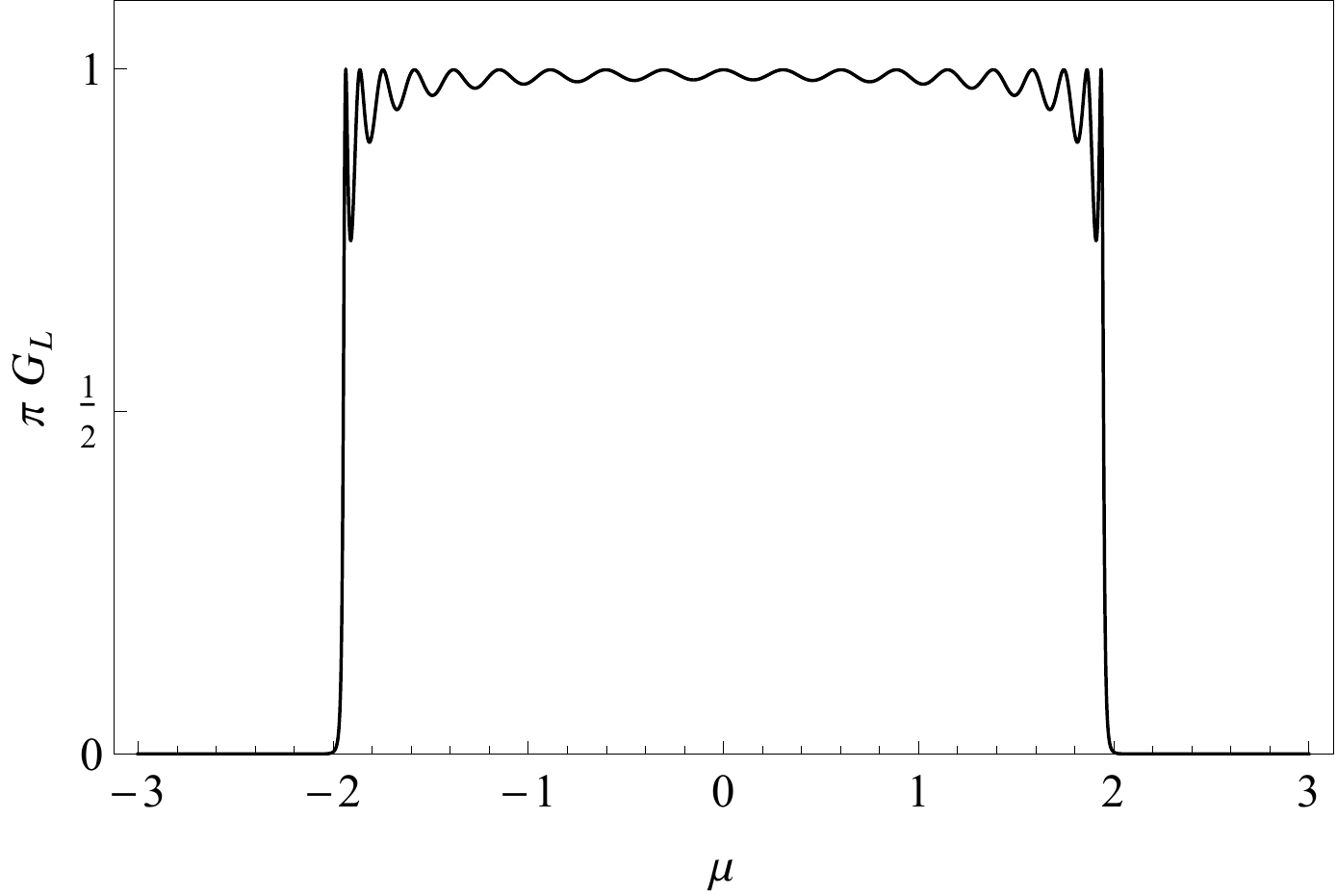}\\
\caption{Conductance of the Kitaev chain at $\ss = 0.4$ and $\D = 0.2$, for $N = 19$.}
\label{fig:eig2}
\end{figure}

To understand the differences between an even or an odd number of sites better it is enlightening to plot the conductance as a function of \m, for a fixed value of $\ss$. In this section, we choose $\ss = 0.4$. For very small values of $\D$, we find (Fabry-P\'erot) resonant peaks in the conductivity at specific values of $\mu$, which turn out to correspond to the eigenvalues of the canonical Hamiltonian with $\D=0$. These can be explained physically by the fact that an electron can only tunnel into the chain if there is an energy level available to accommodate it. Moreover, a zero eigenvalue of the canonical Hamiltonian with $\D=0$ exists only for an odd number of sites and not for an even number of sites. In the latter case tunneling into the chain is only possible because of broadening of the density of states in the chain and the conductance therefore becomes nonuniversal and strongly dependent on $\ss$. In Figs.~\ref{fig:eig} and \ref{fig:eig2} we compare typical plots of the conductance at $\D = 0$ and $\D = 0.2$. We find that the highly oscillatory behavior is almost completely absent when $\D$ is set to a nonzero value. This can be explained by the fact that we are now not probing bulk states, but instead (Majorana) edge states. In addition, the peak conductance is doubled from ${1}/{2\pi}$ to ${1}/{\pi}$. The latter is due to the fact that in the normal state the conductance on resonance is due to two equal contact resistances in series, whereas in the superconducting case effectively only a single contact resistance is present as we have seen before.

It turns out that for a given value of $\ss$, and if $N$ is made large enough, the conductance approaches the perfect `box' shape as a function of $\mu$ for arbitrarily small $\D > 0$. This is consistent with the picture obtained from Sec.~\ref{sec:generald}. From our numerical investigations it appears that the value of $\D$ required to obtain an approximate box shape is inversely proportional to $N$. We can also derive this analytically as follows. We first note that Eq.~(\ref{eqn:deven}) can be rewritten as
\al{
\si_L = \frac{1}{\pi} \lbb 1 + \frac{ 16(1-\D^2)^N + \ss^4 (1-\D^2)^{N-2}}{ 8\ss^2 P_{2N-2} } \rbb^{-1}.
}
For simplicity, take $\ss$ small enough such that the $\ss^4$ term can be neglected. The condition that $G_L$ approaches ${1}/{\pi}$ is then equivalent to the outer fraction being close to $1$. If $N$ is larger than about 10, this will already hold for small $\D$. Expanding to first order in $\D$ gives then the condition $\D \gg 2 \lvert \log{\ss} \rvert/N$, confirming the $1/N$ dependence.

\subsection{Local current}
\label{sec:local}

\begin{figure}[H]
\includegraphics[width=230px]{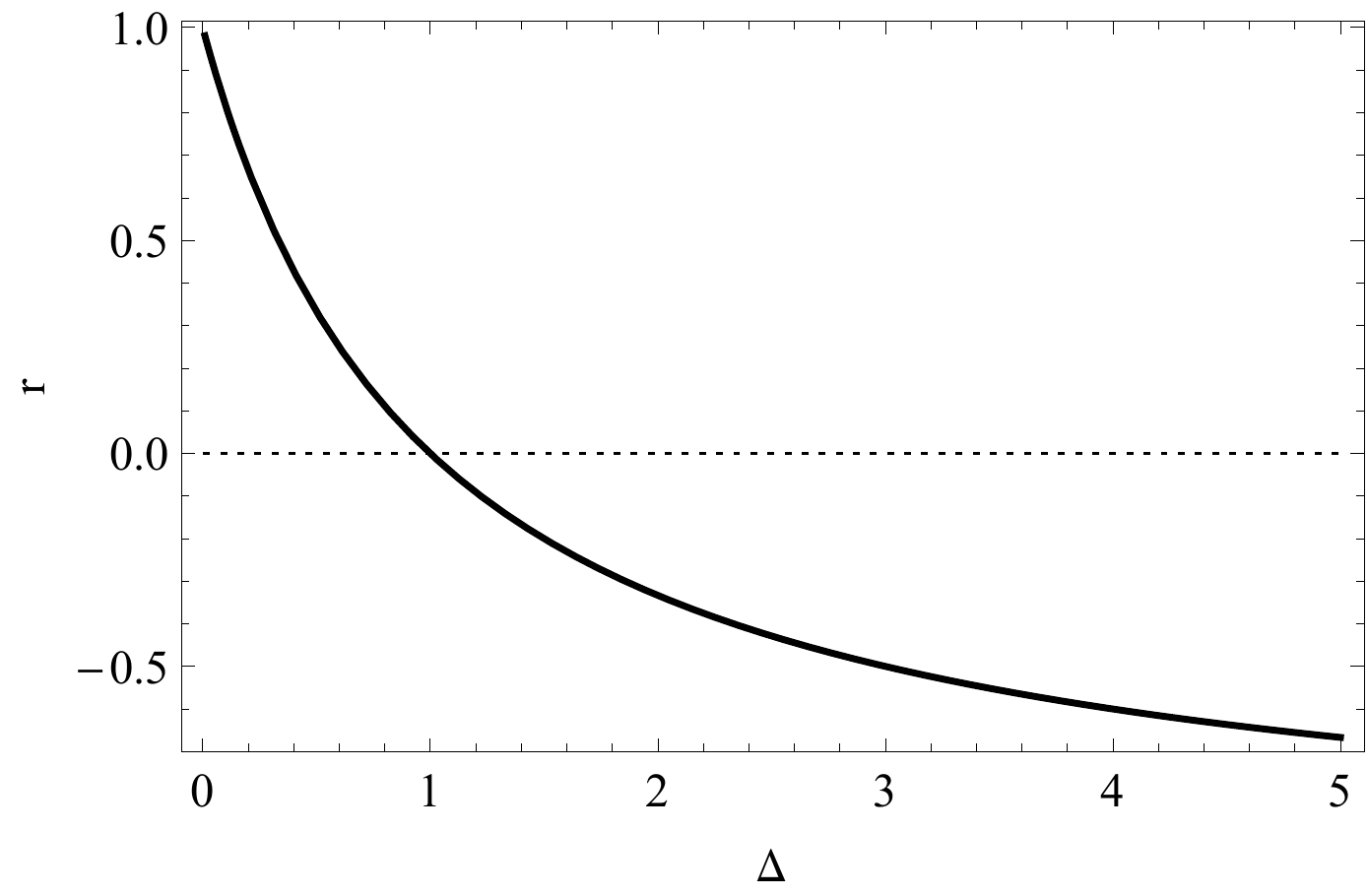}\\
\caption{The ratio $r$ of the local currents as a function of $\D$. The values in this graph turn out to be essentially universal.}
\label{fig:localdecay}
\end{figure}

We have seen in Sec. \ref{sec:delta0} how the electron transport in the exactly solvable Kitaev chain with $\D = 1$ is confined to the boundaries. We now generalize this discussion to other values of $\D$. Plotting the local conductance $\si_{j,j+1,L}$ as a function of the site index $j$, we find that it decays exponentially to an excellent degree of approximation, with a decay length depending on $\D$. For $\D > 1$, the local conductance becomes an alternating function of $j$ but also with an exponentially decaying envelope. In general, we can therefore write $\si_{j,j+1,L}/\si_{j-1,j,L} \equiv r$, where the constant $r$ is negative for $\D > 1$. The ratio $r$ is plotted in Fig.~\ref{fig:localdecay}. It turns out that $r$ is essentially independent of $\ss$, and is given to an excellent degree of approximation by
\be
r(\D) = \frac{1-\D}{1+\D}~.
\ee
We believe that this expression is exact in the thermodynamic limit $N \rightarrow \infty$. Already for $N = 10$ the relative error is less than $10^{-6}$ for typical values of $\D$ in the range $(1/2,2)$. In order to determine $r$ accurately, we measured the local conductance only in the first half of the chain as this reduces distortion due to the other end. As expected, $r$ vanishes exactly at $\D = 1$. This signifies the fact that all local currents vanish except at the endpoints, as noted already in Sec. \ref{sec:delta0}.

\begin{figure}[H]
\includegraphics[width=230px]{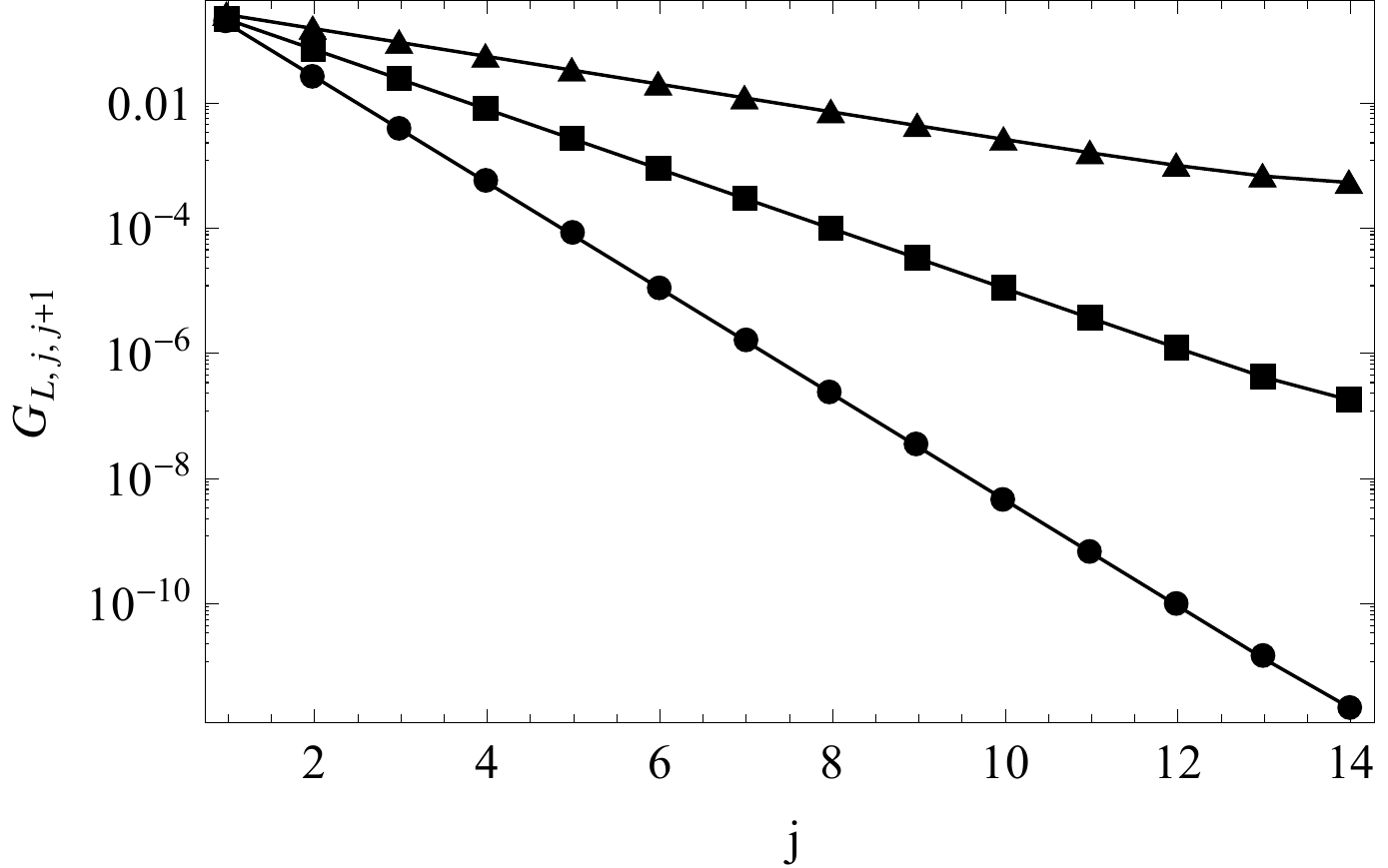}\\
\caption{Local conductance of the Kitaev chain for $\ss = 2$, $N = 15$, and $\D = 1/4$ (triangles), $\D = 1/2$ (squares), and $\D = 3/4$ (disks). A slight deviation from the perfect exponential decay is found only very close to the reservoir on the right of the chain.}
\label{fig:local}
\end{figure}

\begin{figure}[H]
\includegraphics[width=230px]{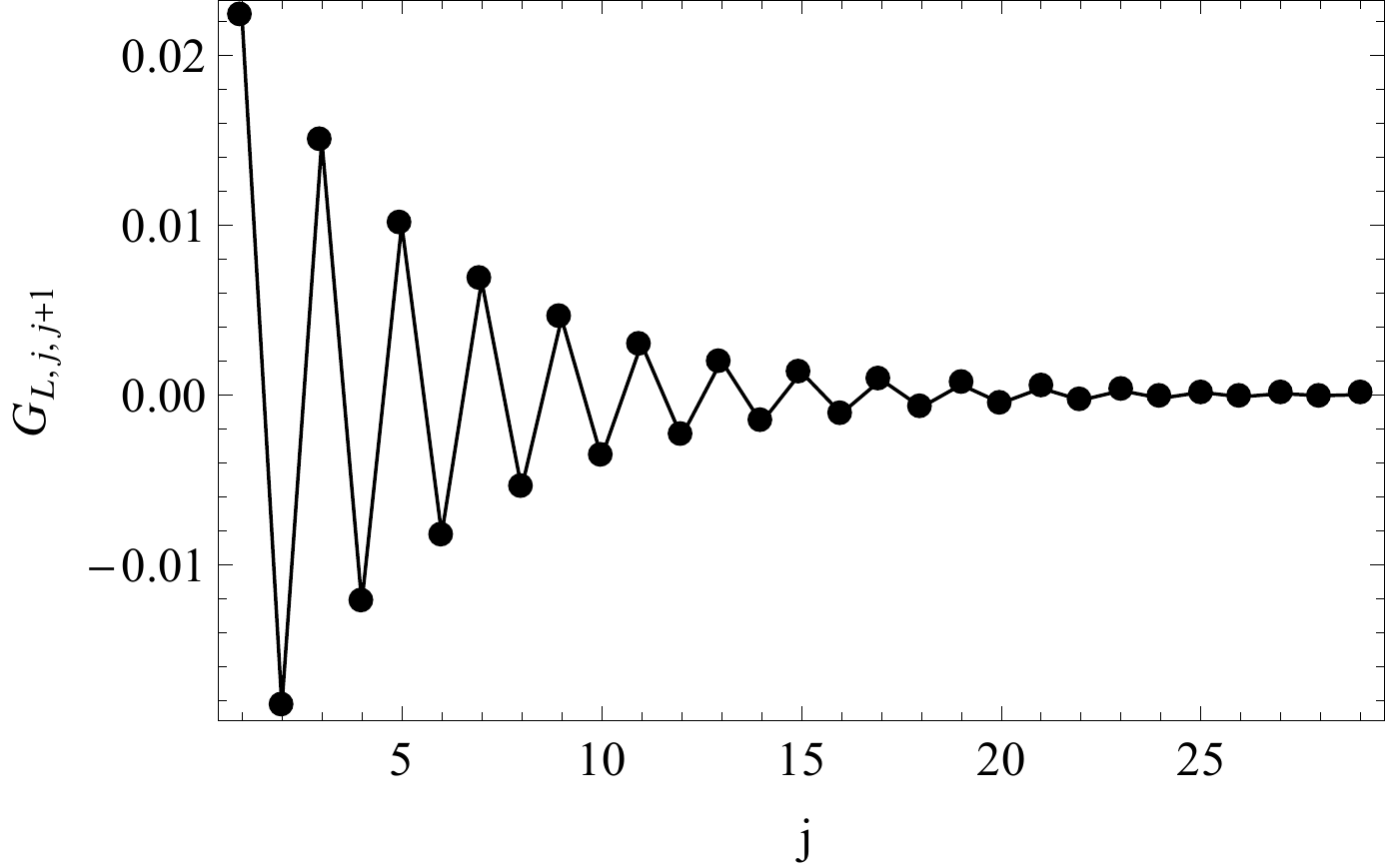}\\
\caption{Local conductance of the Kitaev chain for $\ss = 2$, $N = 30$, and $\D = 10$. The large value of $\D$ increases the decay length, allowing us to see the exponential decay on a linear plot.}
\label{fig:local2}
\end{figure}

\subsection{Current-voltage curves}

\begin{figure}[H]
\includegraphics[width=230px]{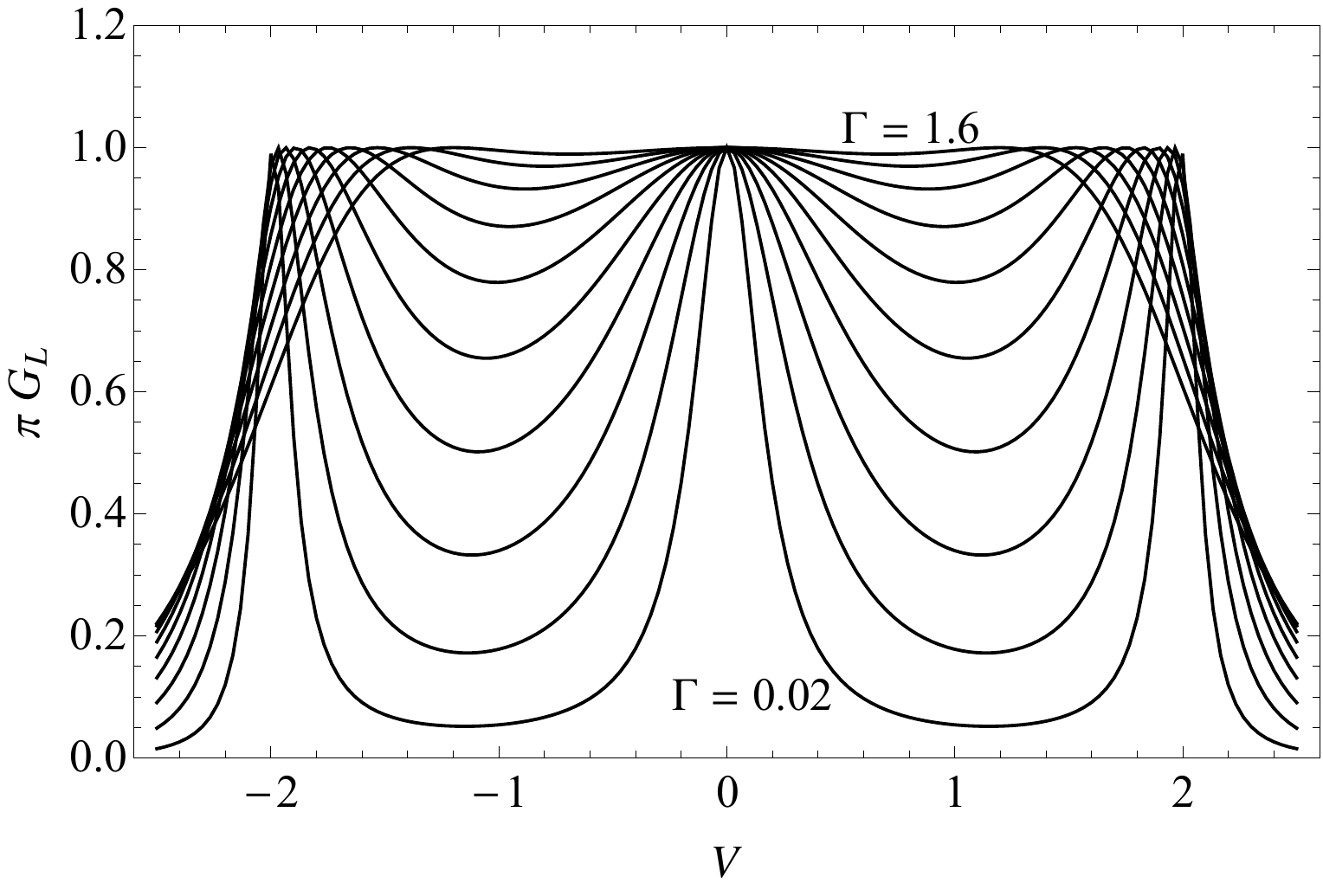}\\
\caption{Differential conductance of the Kitaev chain as a function of applied voltage for various values of $\ss$, with $\m = 0$, $\D = 1$, and $N$ arbitrary.}
\label{fig:diffcon}
\end{figure}

\begin{figure}[H]
\includegraphics[width=230px]{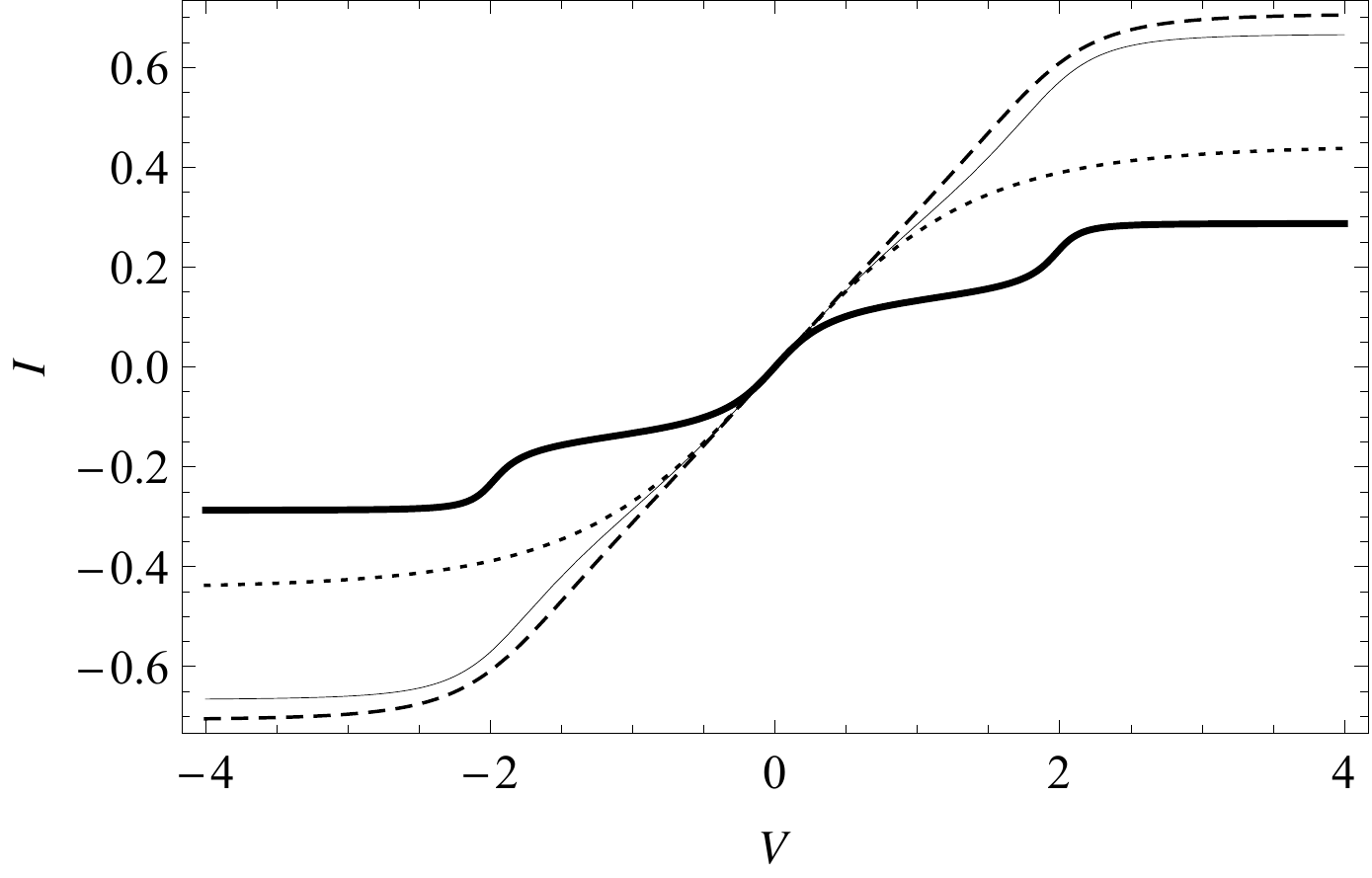}\\
\caption{Full $I$-$V$ curves of the Kitaev chain for $\ss = 0.6$ (thick solid), $\ss = 2$ (thin solid), $\ss = 2 \sqrt 2$ (dashed), $\ss = 8$ (dotted), with $\m = 0$, $\D = 1$, and $N$ arbitrary.}
\label{fig:IV}
\end{figure}

In this section we take $\m_R = \m$, but go beyond linear response theory by allowing $V \equiv \m_L - \m$ and $\ssr$ to be nonzero. For general values of $\m$ and $\D$, the expressions for the conductance become rather complicated and we therefore do not reproduce them here. However, for the particle-hole symmetric case $\m = 0$ and $\D = 1$ we find, denoting by $I$ the current flowing out of the left reservoir, that the differential conductance obeys
\alll{
\frac{dI}{dV} = \frac{4 \ss^2}
{\pi
\Big[ \ss^2/4 + (V - \ssr)^2 \Big]
\Big[ V^2 \ss^2/4 + \lbb 4 - V(V-\ssr) \rbb^2 \Big]}
}
for arbitrary values of $N$. In the remainder we will set $\ssr = 0$ again. The resulting function is shown in Fig.~\ref{fig:diffcon} and can be integrated analytically to obtain the current at any voltage $V$. We do not reproduce the complicated result here, but instead show the $I$-$V$ curves for various values of $\ss$ in Fig.~\ref{fig:IV}.

\begin{figure}[H]
\includegraphics[width=230px]{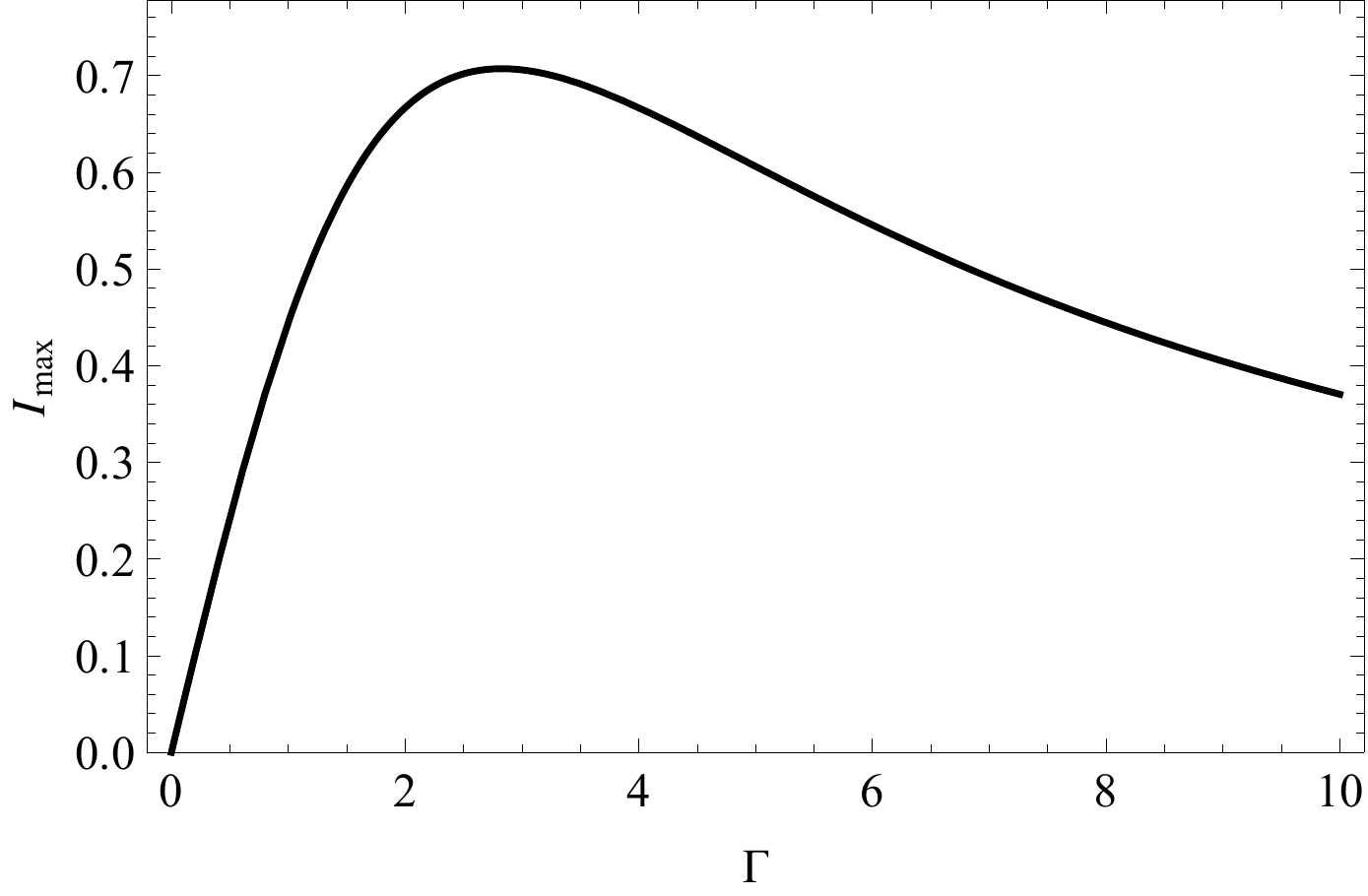}\\
\caption{Limiting value of the current in the Kitaev chain as a function of $\ss$, with $\m = 0$, $\D = 1$, and $N$ arbitrary.}
\label{fig:Imax}
\end{figure}

For sufficiently small values of $\ss$, we find that the current increases significantly around $V = 2$. This corresponds to a local maximum in the differential conductance, which becomes sharper as $\ss$ becomes smaller. The limiting function for small $\ss$ and $V>0$ appears to be $I(V) = \ss \lb 1 + \theta(V - 2) \rb/4$, where $\theta(x)$ denotes the Heaviside function. This is consistent with the fact that the bulk spectrum is dispersionless with a gap equal to 2 in this case \cite{Alicea2012}. It is interesting that the limiting value of the current for $V \rightarrow \infty$ depends strongly on $\ss$, exhibiting a maximum of $1/\sqrt{2}$ at $\ss = 2\sqrt{2}$ as shown in Fig.~\ref{fig:Imax}. The linear-response regime is also found to be maximal near this value of $\ss$.

\section{Conclusions}

In this paper, we have used a stochastic formulation of the Keldysh theory to calculate the lesser and greater Green's functions of the Kitaev chain. We use this method to calculate the conductance of a finite Kitaev chain coupled to two electron reservoirs. This yields exact expressions in special cases. We study the dependence of the conductance on the number of sites, and find that only in the thermodynamic limit the conductance in the topological superconducting regime of the Kitaev model tends to the universal value of $2e^2/h$. The current in the linear-response regime is found to decay exponentially inside the chain. We calculated the decay length and conjectured an exact expression for this length, which is turns out to be very accurate when the system is at least moderately large. Finally, we briefly study the differential conductance and current-voltage curves in the exactly solvable case where the Majorana fermions are perfectly localized. We then also find a nonmonotonic dependence of the maximal current that can flow through the chain on the coupling strength with the reservoir.

Since we focused mostly on obtaining analytical results, all our calculations have been performed at zero temperature. However, extending our work to nonzero temperatures is straightforward. To come closer to experiments it is also interesting to consider chains that are segmented into a number of superconducting regions. This can easily be incorporated into our approach, by appropriately modifying the pairing matrix ${\bf D}$ in Eq.\ (5), and in this manner the chain will contain more than two Majoranas. Of course, ultimately also an additional disorder average needs to be introduced into the model to be able to contribute more to the ongoing discussion \cite{lee2014spin} on the correct interpretation of the exciting experiments with semiconductor nanowires.

\section*{acknowledgements}

We thank Rembert Duine, Lars Fritz, and Dirk Schuricht for useful and stimulating discussions. This work was supported by the Stichting voor Fundamenteel Onderzoek der Materie (FOM), the Netherlands Organization for Scientific Research (NWO) and the Delta-Institute for Theoretical Physics (D-ITP).

\bibliography{FiniteKitaevChain}{}

\end{document}